\documentstyle[aps,prb,epsf,twocolumn,floats,amssymb,amsfonts]{revtex}
\begin{document}
\bibliographystyle{prsty}
\global\firstfigfalse
\global\firsttabfalse
\title{Mesoscopic fluctuations of the ground state spin of a small
 metal particle}
\author{P.\ W.\ Brouwer, Yuval Oreg, and B.\ I.\ Halperin}
\address{ Lyman Laboratory of Physics, Harvard University, Cambridge
  MA 02138 \\ {\rm \today} \\ \bigskip \parbox{14cm} 
  {\rm We study the statistical distribution of the ground state spin
    for an ensemble of small metallic grains, using a random-matrix
    toy model. Using the Hartree Fock approximation, we find that 
    already for interaction strengths well below the Stoner criterion
    there is an appreciable 
    probability that the ground state has a finite, nonzero spin.
    Possible relations to experiments are discussed.
    \smallskip \\ PACS numbers 
73.23.-b, 71.10.-w, 71.24.+q, 75.10.Lp\vspace{-0.5cm}
} } \maketitle

%
%
%
%
%

According to Hund's rule,\cite{LL:QMwf} electrons in a partially filled
shell in an atom form a many-body ground state with maximum possible
spin. The maximum spin is preferred because it allows a maximally
antisymmetric coordinate wavefunction in order to minimize the
electrostatic repulsion between the electrons. In recent
experiments,\cite{WF:Tarucha96} Hund's rule was also observed in a
cylindrically-shaped semiconductor quantum dot, or ``artificial
atom''.  The close similarity with real atoms is due to the degeneracy
of single-particle levels, caused by the the high degree of symmetry of
the device.
 
In generic ultrasmall systems such as small metal
grains,\cite{WF:Ralph95,WF:Davidovic98} semiconductor quantum
dots,\cite{WF:Stewart97,WF:Review} or carbon
nanotubes\cite{WF:Tans98,WF:Cobden98} there is no systematic degeneracy
due to a spherical (or cylindrical) symmetric potential. However, even
in the absence of degeneracies, a nonzero value of the ground state
spin may occur, as long as the gain in electrostatic energy is larger
than the loss in kinetic energy when an antisymmetric coordinate ground
state wavefunction is formed.  Such a ground state is most likely to be
detected in ultrasmall metal and semiconductor devices, since in those
systems, unlike in macroscopic samples, the spacing between
single-particle energy levels and the typical interaction energies can
be larger than the temperature.  In fact, the possibility of such a
``weakly ferromagnetic'' ground state has been suggested as an
explanation for some recent experiments, that could not be explained by
simple noninteracting models.\cite{WF:Davidovic98,WF:Tans98,WF:Prus96}
In addition, a nonzero ground state spin from numerical simulations of
a few particles in a chaotic dot,\cite{WF:Berkovits98} and a theory of
spin polarization in larger dots\cite{WF:Andreev98} were already
mentioned in the literature. The stability of the zero spin ground state 
in a quantum dot was analyzed for weak
interactions in Ref.\ \onlinecite{WF:Prus96}.

In this paper, we consider small metal grains in the mesoscopic regime,
in which fluctuations of wavefunctions and energy levels, caused by,
e.~g., disorder or an irregular shape, control the behavior of kinetic
and interaction energies at the vicinity of the Fermi energy.  As a
result, the ground state spin becomes subject to sample-to-sample
fluctuations. Then, the relevant quantity to consider is the
statistical distribution of the ground state spin for an ensemble of
small metal grains or chaotic quantum dots, rather than the spin of a
specific sample.  

Our starting point is
a simple toy model that captures the essential
mechanisms for mesoscopic fluctuations of the ground state spin. In
second-quantized form, our model Hamiltonian ${\cal H}$ reads
\begin{eqnarray} \label{eq:Ham}
  {\cal H} &=& \sum_{n,m,\sigma}  
    c^{\dagger}_{n,\sigma} {\cal H}_0(n,m) 
      c^{\vphantom{\dagger}}_{m,\sigma} \nonumber \\ && \mbox{}
    + u M \sum_{n} c^{\dagger}_{n,\uparrow} c^{\dagger}_{n,\downarrow}
      c^{\vphantom{\dagger}}_{n,\downarrow}
      c^{\vphantom{\dagger}}_{n,\uparrow},
\end{eqnarray}
where $c^{\dagger}_{n,\sigma}$ ($c^{\vphantom{\dagger}}_{n,\sigma}$)
is the creation (annihilation) operator for an electron with spin
$\sigma$ at site $n$. The indices $m,n$ are summed over $M$
sites.  The Hamiltonian ${\cal H}_0$ contains the kinetic energy and
the impurity potential. We describe the electron-electron interaction
by an on-site (Hubbard) interaction, $uM$. While the long-range
Coulomb interaction can be trivially included via a charging
energy,\cite{WF:Review} the model (\ref{eq:Ham}) does not include the
Coulomb interaction at intermediate distances, which leads to a
Gaussian level spacing distribution
at the Fermi energy.\cite{WF:HF} 
In this work, we report a calculation of the ground state spin of the
Hamiltonian (\ref{eq:Ham}) using a restricted version of the
Hartree-Fock (HF) approximation with a random-matrix assumption for the
eigenvalues and eigenvectors of the self-consistent HF Hamiltonian.

We first present our main result. It consists of an equation that
relates the candidate ground state energies $E_{\rm G}(s)$ for
different values of the total spin $s$ in terms of eigenvalues
$\varepsilon_{\mu}^{0}$ of a hermitian random matrix with level
spacing $\Delta$, the interaction parameter $\lambda = u/\Delta$, and
a (nonuniversal) numerical constant $c$ that describes the density
response to a local perturbation of the impurity potential in ${\cal
  H}_0$,
\begin{eqnarray}
  E_{\rm G}(s) - E_{\rm G}(s_0) &=& 
  \sum_{\mu=1}^{s} 
  (\varepsilon_{N+\mu+2 s_0}^{0} - \varepsilon_{N+1-\mu}^{0}) \nonumber \\
  && \mbox{} - 
  \lambda \Delta \left[ s^2 - s_0^2 + {2 (s - s_0) \over 
    \beta(1 - \lambda^2 c^2)} \right].
  \label{eq:EGdiff}
\end{eqnarray}
The total number of electrons is $2(N+s_0)$, $s_0$ being $0$ or $1/2$.
The spin of the true ground state is found by minimizing Eq.\ 
(\ref{eq:EGdiff}) with respect to $s$.  The parameter $\beta=1$ ($2$)
if time-reversal symmetry is present (absent). The effect of spin
orbit coupling and Zeeman splitting is not included here.
(The case $\beta=2$ is only relevant for semiconductor quantum dots in
a weak magnetic field, that affects orbital motion, but causes no 
Zeeman splitting. It is not relevant for 
small metal grains,\cite{WF:Davidovic98} as
laboratory magnetic fields do not affect orbital motion in this case.) 
Equation (\ref{eq:EGdiff}) reflects the competition between kinetic
energy (first term on the r.h.s.), which favors small $s$ and the
on-site interaction (second term), which favors finite $s$. The
interaction term, in turn, consists of two parts: A term quadratic in
$s$, which describes the exchange interaction,
and a
term linear in $s$, which describes the additional ``dressed'' Coulomb
repulsion of two particles with the same spatial 
wavefunction.
For large $s$, the contribution from the kinetic term is approximately
$s^2 \Delta$, so that for $u \gtrsim \Delta$ a finite fraction of the
total number of spins will align, rather than a small number of spins
as in the case $u < \Delta$. The instability at $u = \Delta$ is known
as the Stoner instability. 
In Ref.\ \onlinecite{WF:Prus96}, a result similar to
Eq.\ (\ref{eq:EGdiff}) was obtained for $s=1$, but with a different
and fluctuating interaction term. The difference is due to the absence of
a self-consistent approximation scheme in Ref.~\onlinecite{WF:Prus96}.

\begin{figure}
  \vglue -2.2cm \epsfxsize=0.95\hsize
  \epsffile{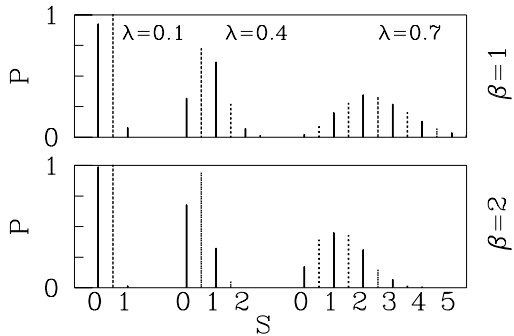}\vspace{-1.1cm}

\caption{\label{fg:Ps} The probability distribution $P(s)$ of the
   ground state spin of a small metal grain, computed from Eq.\
   (\protect\ref{eq:EGdiff}) for three different values of the 
   interaction parameter $\lambda$. The upper (lower) histograms are for
   the presence (absence) of time-reversal symmetry. Solid histograms are
   for integer spin, dotted ones for half-integer 
   spin. (The density-response
   parameter $c$ has been set to zero; finite $c$ results in an even
   higher probability to find nonzero spin.) \vspace{-0.2cm}}
\end{figure}

As a consequence of the additional dressed Coulomb repulsion of particles
with the same wavefunction, we find that already for interaction
strengths considerably below the Stoner instability $u = \Delta$, there is an
appreciable probability of nonzero ground-state spin. This is
illustrated in Fig.\ \ref{fg:Ps}, where the distribution of the ground
state spin at three different values of the interaction parameter
$\lambda$ is shown: Already at the quite modest interaction strength $u
\approx 0.4 \Delta$ a ground state spin $s=1$ is more likely than $s=0$.

The effect of a weak magnetic field is twofold: First, it changes the
statistics of the $\varepsilon_{\mu}^0$,\cite{RMT:Mehta91} and, second,
it suppresses the interference in the ``Cooper channel'', leading to a
factor of two reduction of the interaction term linear in $s$ [last
term in Eq.\ (\ref{eq:EGdiff})]. Both effects favor lower $s$ than
without a magnetic field.  However, even in the absence of a magnetic
field we expect that, similar to 3D metals,\cite{SC:Morel62} inclusion
of the electron--electron interaction beyond the HF approximation will
also lead to a suppression of the interference in the Cooper channel
(logarithmically in the system size $M$), and hence to a prefactor in
that term that is smaller than 2.

Let us now turn to the details of our calculation. To find the ground
state of the Hamiltonian (\ref{eq:Ham}) we use a simplified version of
the HF approximation: We assume that the ground state has the form of a
Slater determinant of single-particle wave functions
$\psi_{\mu,\uparrow}$ and $\psi_{\mu,\downarrow}$ of particles which
have either spin up or spin down.  In this case the self-consistent HF
equations read
\begin{eqnarray} \label{eq:HHF}
  {\cal H}_{{\rm HF},\sigma} \psi_{\mu,\sigma}& =& 
    \varepsilon_{\mu,\sigma} \psi_{\mu,\sigma};\nonumber \\
  {\cal H}_{{\rm HF},\sigma}(n,m) &=& 
    {\cal H}_0(n,m) + u M \rho_{-\sigma}(n) \delta(n,m), \\
  \rho_{\sigma}(n) &=&
    \sum_{\mu} f_{\mu,\sigma} |\psi_{\mu}(n)|^2 \nonumber.
\end{eqnarray}
The occupation number $f_{\mu,\sigma}$ is $1$ ($0$) if the level
$\mu,\sigma$ is occupied (unoccupied) and $\delta(n,m)$ is the
Kronecker delta function. The ground state energy $E_{\rm G}$ is given
by
\begin{equation} \label{eq:EG}
  E_{\rm G} = 
    \sum_{\mu,\sigma} f_{\mu,\sigma} \varepsilon_{\mu,\sigma} -
    u M \sum_{n} \rho_{\downarrow}(n) \rho_{\uparrow}(n).
\end{equation}

Our strategy is as follows: We start from a reference state with zero
spin, in which $N$ particles of each spin are placed in the same levels
$\varepsilon_{\mu,\uparrow} = \varepsilon_{\mu,\downarrow}$ and with
the same wavefunctions $\psi_{\mu,\uparrow}=\psi_{\mu,\downarrow}$. We
assume, that for this symmetric case the eigenvectors and the
eigenvalues of ${\cal H}_{\rm HF}$ are distributed like those of a
random matrix, except that the energy levels below $E_F$ are shifted
upwards, by a small constant amount, relative to the levels above
$E_F$, see Eq.\ (\ref{eq:Emu}) below. (If we would have included
long-range Coulomb interactions via a charging energy, the shift would
have been much larger and in the opposite direction.
Omission of the charging energy has no consequence in our case, as 
we compare ground states with the same number of particles.) 
The assumption that the single-particle eigenvalues and wavefunctions
in a self-consistent potential for a mesoscopic system below and
above $E_F$ obey random
matrix statistics, even though they may be quite different from their
counterparts in the noninteracting system,
was checked numerically for short range interaction models somewhat
similar to ours.\cite{WF:HF}
The energy shifts in our case result from the spin-degeneracy,
which was not present in these calculations.

Starting from this reference state, we build other states by the
subsequent addition and removal of electrons.
We first discuss the addition of a single up spin in the $(N+1)$st
level. The first question that needs to be answered is how this
addition affects the self-consistent density $\rho_{\sigma}(n)$.  The
density change $\delta \rho_{\uparrow}(n)$ consists of a direct and an
induced contribution, while $\delta \rho_{\downarrow}(n)$ has an
induced density shift only,
\begin{eqnarray}
  \delta \rho_{\uparrow} &=& 
    \delta \rho_{\uparrow,{\rm dir}} + \delta \rho_{\uparrow,{\rm ind}}, 
    \ \ 
  \delta \rho_{\uparrow,{\rm dir}}(n) = |\psi_{N+1,\uparrow}(n)|^2,
    \nonumber \\ 
  \delta \rho_{\downarrow} &=& 
    \delta \rho_{\downarrow,{\rm ind}}. \label{eq:rhodefs}
\end{eqnarray}
Since the density shifts change the HF Hamiltonians by an
amount 
$\delta {\cal H}_{{\rm HF},\sigma}(n,m) = u M \delta \rho_{-\sigma}(n)
\delta(n,m)$,
we obtain the following self-consistency equations for $\delta \rho_{\sigma}$,
\begin{eqnarray}
  \delta\rho_{\sigma,{\rm ind}}(n) &=&
  2\, uM\, \mbox{Re}\, 
    \sum_{\mu, \nu, m} f_{\mu,\sigma} (1-f_{\nu,\sigma})
  \delta \rho_{-\sigma}(m) \nonumber \\ && \mbox{} \times
  {\psi_{\mu,\sigma}^{*}(n)
   \psi_{\mu,\sigma}^{\vphantom{*}}(m)
   \psi_{\nu,\sigma}^{*}(m) 
   \psi_{\nu,\sigma}^{\vphantom{*}}(n)
    \over
   \varepsilon_{\mu,\sigma} - \varepsilon_{\nu,\sigma}}.
   \label{eq:deltarho1}
\end{eqnarray}
Both $\delta \rho_{\uparrow}$ and $\delta \rho_{\downarrow}$ are of
order $1/M$.  In Eq.\ (\ref{eq:deltarho1}) we have computed the induced
density change to first order in $\delta {\cal H}_{{\rm HF}}$. Higher
order terms do not contribute to $\delta \rho_{{\rm ind},\sigma}$ to
order $1/M$ and are neglected.
To evaluate Eq.\ (\ref{eq:deltarho1}), we first sum the r.h.s.\
over the space
index $m$ and then over the energy levels $\varepsilon_{\mu}$ and
$\varepsilon_{\nu}$.
Because the eigenfunction has a random sign, a single term in the
latter summation is of order
$\lambda M^{-3/2} \Delta/(\varepsilon_{\mu} - \varepsilon_{\nu})$,
which is not relevant in the limit $M \gg 1$, even if
$\varepsilon_{\mu}$ and $\varepsilon_{\nu}$ are both close to the
Fermi level $E_F$. For the summation over all levels we may perform an
average over the wave functions (since the denominator is a slowly
varying function of $\mu$ and $\nu$ away from the Fermi level).  This
average is done using that for general $\mu \neq \nu$ and
in the limit $M \gg 1$ one has,\cite{RMT:Mehta91}
$$
  \langle \psi_{\mu,\sigma}^{*}(n)
   \psi_{\mu,\sigma}^{\vphantom{*}}(m)
   \psi_{\nu,\sigma}^{*}(m) 
   \psi_{\nu,\sigma}^{\vphantom{*}}(n) \rangle
  = {\delta(m,n) \over M^2} - {1 \over M^3}.
$$

Putting everything
together, we find the following solution of the self-consistency
equations (\ref{eq:rhodefs}) and (\ref{eq:deltarho1}),
\begin{eqnarray}
  \delta \rho_{\uparrow}(n) &=& {1 \over 1 - (c \lambda)^2} \left( |\psi_{N+1,\uparrow}(n)|^2 - {1 \over M} \right) + {1 \over M}, \nonumber \\
  \delta \rho_{\downarrow}(n) &=& {- c \lambda \over 1 - (c \lambda)^2} \left( |\psi_{N+1,\uparrow}(n)|^2 - {1 \over M} \right). \label{eq:dRho}
\end{eqnarray}
where $c$ is a numerical constant of order unity defined by
\begin{equation}
  c = \lim_{M \to \infty}
  {2 \Delta \over M} \int_{-\infty}^{E_F} d \varepsilon_1 
  \int_{E_F}^{\infty} d \varepsilon_2 {\rho(\varepsilon_1) \rho(\varepsilon_2) \over \varepsilon_2 - \varepsilon_1}.
\end{equation}
In this equation,
$\rho(\varepsilon)$ is the mean density of HF energy levels. [The mean
level spacing $\Delta$ is taken at the Fermi energy, $\Delta
= 1/\rho(\varepsilon_F)$.]
The constant $c$ gives the linear density response $\delta \rho(n)$ to
a shift of the impurity potential ${\cal H}_0(n,n)$ at that same site,
$\delta \rho(n) = c\, \delta {\cal H}_0(n,n)/M \Delta$, which can be 
verified using
first order perturbation theory with respect to $\delta {\cal H}_0$. 
Notice that
$c$ is not a universal constant, but depends on an integration of the
density of states over the entire bandwidth. For example, for the Wigner
semicircular density of states we find $c=4/3$ if the Fermi energy
$E_F$ is at the band center and $c \to 0$ if $E_F$ is at a 
band edge. 
Equation (\ref{eq:dRho}) expresses that the interaction enhances
the fluctuations of the spin density: if $|\psi_{N+1,\uparrow}(n)|^2$
is larger than average, the on-site repulsion reduces
$\rho_{\downarrow}(n)$, which in turn causes an increase of
$\rho_{\uparrow}(n)$, and so on. At the same time, the interaction
reduces fluctuations of the charge density $\rho_{\uparrow} + 
\rho_{\downarrow}$.

At $\lambda c = 1$, which may occur before the Stoner instability
$\lambda = 1$ if $c > 1$, the density changes diverge. Although this
instability signals a breakdown of our approach, it is not clear
whether it will also lead to a true macroscopic ground state spin.
Below, we restrict our discussion to the case $\lambda c < 1$.

Next we address the 
HF energy levels $\varepsilon_{\mu,\sigma}$ and find\vspace{-0.3cm}
\begin{eqnarray}
\delta \varepsilon_{\mu,\uparrow}&= & -{\lambda \Delta} 
  {2 \over \beta} { c \lambda \over 
    1 -(c \lambda) ^2}\, \delta_{\mu,N+1},  \nonumber \\
  \delta \varepsilon_{\mu,\downarrow}&=& 
    \lambda \Delta \left( 1 + {2 \over \beta}
    {1 \over 1 - (c \lambda)^2}\, \delta_{\mu,N+1} \right) 
 \label{eq:dE} .
\end{eqnarray}
\vspace{-0.4cm}

\noindent
The shift of $\varepsilon_{N+1,\sigma}$ is extra large, since
for that level the interaction effects are enhanced by the spatial
fluctuations of the wavefunction.  Equation (\ref{eq:dE}) is the
result of first order perturbation theory in $\delta \rho_{\sigma}$;
second order perturbation theory $\delta
\varepsilon_{\mu,\sigma}$ gives a correction of order $\lambda^2
\Delta \ln M/M$, which we may neglect in the limit $M \gg 1$. In the
same way, one finds that the changes in each individual wave functions
is not significant for $M \gg 1$.

Finally, we consider the change in the ground state energy $E_{\rm G}$. 
Because the summation over $\mu$ in
Eq.\ (\ref{eq:EG}) extends over ${\cal O}(M)$ levels, it is important
to follow the shifts in the HF levels to second order
perturbation theory, although this level of accuracy was not needed
for the shift of each
level individually, cf.\ Eq.\ (\ref{eq:dE}). 
Putting everything together, we find that
\begin{equation} \label{eq:dEG}
 \delta E_{\rm G} = \varepsilon_{N+1,\uparrow}   - 
c \lambda^2 \Delta [\beta(1-c^2\lambda^2)]^{-1}.
\end{equation}
(No terms proportional to $\log M$ appear here since they cancel in the
summation over the energy levels.) With the help of Eq.\ (\ref{eq:dE}),
we can rewrite Eq.\ (\ref{eq:dEG}) as $\delta E_{\rm G} =
\varepsilon_{N+1,\uparrow} + {1 \over 2} \delta
\varepsilon_{N+1,\uparrow}$, which is the average of the energy for
the newly occupied level $\varepsilon_{N+1,\uparrow}$
before and after its occupation.
This may be interpreted as a simple extension of
Koopmans' theorem \cite{WF:Koopmans34} to the present case, where the
modification in each one-electron wavefunction is small (of relative
order $M^{-1/2}$), but the resulting contribution to $\delta E_{\rm G}$
cannot be neglected to the order we are interested in.  In the usual
form of Koopmans' theorem, where one ignores any change in the
one-particle wavefunctions, the HF energies of the lowest unoccupied
state before addition of the electron, and of the highest occupied
state after addition are identical.  (The usual Koopmans' theorem is
correct for an infinite system, in general, or for a translationally
invariant finite system, as the one-electron states are trivially plane
waves in that case.)  The simple extension of Koopmans' theorem also
works, in our model, for the addition of several electrons.

We have repeated these calculations for the addition of two electrons
with opposite spin in the $(N+1)$th level,\vspace{-0.2cm}
\begin{eqnarray}
  \delta \rho_{\sigma}(n) &=& 
    {1 \over M} + 
    {1 \over 1 + c \lambda} \left( |\psi_{N+1,\uparrow}(n)|^2 -
      {1 \over M} \right), \\
  \delta \varepsilon_{\mu,\sigma} &=&
    \lambda \Delta
    \left(1 +  {2 \over \beta} {1 \over 1 + c \lambda }\,
      \delta_{\mu,N+1} \right), \label{eq:dE2} \\
  \delta E_{\rm G} &=&
    2 \varepsilon_{N+1} +
      {2 \over \beta} \lambda \Delta {1 \over 1 + c \lambda}.
\end{eqnarray}
\vspace{-0.3cm}

\noindent
As in the case of the addition of a single particle, the individual
wavefunctions do not change to order $M^{-1/2}$.  Equation
(\ref{eq:dE2}) allows us to find the statistics of the HF energy levels
$\varepsilon_{\mu,\sigma}$ in our reference system with $N$ electrons
of each spin:  The only distribution that is consistent both with the
assumption that the $\varepsilon_{\mu,\sigma}$ obey random-matrix
statistics away from the Fermi level and with the shifts of
Eq.\ (\ref{eq:dE2}) is one where the $\varepsilon_{\mu,\sigma}$ have
the form\vspace{-0.2cm}
\begin{equation} \label{eq:Emu}
  \varepsilon_{\mu,\sigma} = \varepsilon_{\mu}^{0} + {2 \over \beta} f_{\mu}
  {\lambda \Delta \over 1+\lambda c},
\end{equation}\vspace{-0.3cm}

\noindent
where the $\varepsilon_{\mu}^{0}$ have random matrix statistics and
$f_{\mu} = 1$ ($0$) if the level $\mu$ is (un)occupied. In other words,
the distribution of the $\varepsilon_{\mu}$ is the same as that of the
eigenvalues of a random matrix, with all occupied levels shifted
upwards by an amount $(2/\beta) \lambda \Delta/(1 + c \lambda)$.

With the knowledge we have gained above, there is little work left for
the calculation of our main result, Eq.\ (\ref{eq:EGdiff}).
Some remarks about the validity of this result are appropriate.  First,
to make a connection between our random matrix toy model and a
laboratory made quantum dot we must identify $M= (L/\lambda_F)^3$ as we
expect the length scale for wavefunction
correlations\cite{WF:Prigodin95} and the range of the screened Coulomb
interaction to be of order of the Fermi wavelength
$\lambda_F$.\cite{WF:RMTfootnote} Second, while our solution is
complete within the HF approximation, one must bear in mind that this
approximation scheme does not include correlation effects, such as the
Cooper channel renormalization, as we discussed below Eq.\
(\ref{eq:EGdiff}).  Those correlation effects are not expected to
affect our result to first order in $\lambda$, but it can not be
excluded that they are important in the higher order terms in Eq.\
(\ref{eq:EGdiff}), which involve the factor $c \lambda$.

We close this paper with a discussion of the physical consequences
of a ground state with spin $s>1/2$ and of the experimental
situations in which it can be observed.

First, the temperature $T$ needs to be smaller than the separation of
the ground states for different spin $s$. This separation, which is
typically smaller than the single particle level spacing $\Delta$, is a
fluctuating quantity. Very small values are possible, because, unlike
in noninteracting random systems, there is no level repulsion if states
of different spin are involved.

For sufficiently low $T$, the magnetization of the grain is
proportional to the spin $s$ of the ground state. However, $s$ will
also affect other properties which are more easily accessible in an
experiment, like current--voltage characteristics:
A nonzero ground state spin can serve to explain the absence of an 
even-odd structure in the addition spectra of Coulomb-blockaded quantum
dots,\cite{WF:Prus96,WF:Berkovits98,WF:Sivan96,WF:Blanter97}
or the presence of kinks in the parametric dependence of Coulomb
blockade peak positions, as was noted in Ref.\ \onlinecite{WF:Baranger99}.
Spin is also relevant for conductance measurements at a finite bias
voltage, which allow for a ``spectroscopy'' of the quantum dot or metal
grain.\cite{WF:Review,WF:Agam97footnote}
In the presence of a magnetic field, the ground state is split by
the Zeeman energy, and the differential conductance will show two
peaks, whose relative intensity differs by a factor $2 s_{N+1} + 1$ or
$2 s_{N} + 1$, depending on whether the tunneling onto or from the
grain is the faster process.\cite{WF:Gueron99} Even without
an external magnetic field the ground state may be split, e.g., by
spin-orbit coupling or magnetic impurities, and thus give rise to a
multiplet of peaks in the differential
conductance.\cite{WF:Davidovic98a} The peak separation within 
a multiplet is controlled by the strength of the splitting mechanism
and may be much smaller than $\Delta$.

We thank D.~Davidovi\'c and M.~Tinkham for stimulating discussions and
for sharing their experimental results with us. When this work was
nearly completed, we learned of related work by H.\ U.\ Baranger,
D.\ Ullmo, and L.\ I.\ Glazman (Ref.\ \onlinecite{WF:Baranger99}), in
which some similar results were obtained. We thank H.\ U.\ Baranger for
discussions on these points.  This work was supported in part by the
NSF through the Harvard MRSEC (grant DMR 98-09363), and by grants DMR
94-16910, DMR 96-30064, DMR 97-14725.


\begin{thebibliography}{10}
\bibitem{WF:RMTfootnote}
  This estimate of $M$ differs from one which is commonly
  used in the noninteracting case where $M$ is chosen to be the
  dimensionless conductance.
  In our case the choice for $M$ is dictated by the range of the
  screened Coulomb interaction.
\end{thebibliography}
\end{document}